\documentclass[a4paper,10pt]{article}

\usepackage[margin=1in]{geometry}

\pdfoutput=1

\usepackage{graphicx}
\usepackage{bm}
\usepackage{dcolumn}
\usepackage{setspace}
\usepackage{abstract}
\usepackage{multirow}
\usepackage{amsmath}
\usepackage{fancyhdr}
\usepackage{cite}
\usepackage{color,soul}

\usepackage[affil-it]{authblk}

\usepackage[utf8]{inputenc}
\usepackage{lineno,hyperref}

\hypersetup{colorlinks=true, linkcolor=blue, citecolor=red}

\begin{document}

\title{\bf Deformation behaviour of body centered cubic iron nanopillars containing coherent twin boundaries}
\date{}

\author{ G. Sainath\footnote{email : sg@igcar.gov.in}, B.K. Choudhary\footnote{email : bkc@igcar.gov.in}}

\affil {Deformation and Damage Modelling Section \\ Mechanical Metallurgy Division
\\ Indira Gandhi Centre for Atomic Research, Kalpakkam \\ 
Tamilnadu-603102, India}

\maketitle

\doublespacing

\begin{onecolabstract}
Molecular dynamics simulations were performed to understand the role of twin boundaries on deformation behaviour 
of body-centred cubic (BCC) iron (Fe) nanopillars. The twin boundaries varying from one to five providing twin boundary 
spacing in the range 8.5 - 2.8 nm were introduced perpendicular to the loading direction. The simulation results 
indicated that the twin boundaries in BCC Fe play a contrasting role during deformation under tensile and compressive 
loadings. During tensile deformation, a large reduction in yield stress was observed in twinned nanopillars compared 
to perfect nanopillar. However, the yield stress exhibited only marginal variation with respect to twin boundary 
spacing. On the contrary, a decrease in yield stress with increase in twin boundary spacing was obtained during 
compressive deformation. This contrasting behaviour originates from difference in operating mechanisms during 
yielding and subsequent plastic deformation. It has been observed that the deformation under tensile loading was 
dominated mainly by twin growth mechanism, due to which the twin boundaries offers a negligible resistance to slip 
of twinning partials. This is reflected in the negligible variation of yield stress as a function of twin boundary 
spacing. On the other hand, the deformation was dominated by nucleation and slip of full dislocations under compressive 
loading. The twin boundaries offer a strong repulsive force on full dislocations resulting in the yield stress 
dependence on twin boundary spacing.  Further, it has been observed that the curved twin boundary can acts as a 
source for full dislocation. The occurrence of twin-twin interaction during tensile deformation and dislocation-twin 
interaction during compressive deformation were presented and discussed. \\

\end{onecolabstract}


{\small

\section{Introduction}

In recent years, the twinned nanopillars or nanowires have drawn growing attention in view of their superior physical 
properties and potential applications in modern small-scale electronic devices such as nano/micro electro mechanical 
systems. The twinned nanopillars contain a series of twin boundaries with specified spacing between the boundaries. 
Twin boundary possesses high symmetry and lowest interface energy along with well-defined boundary plane. The low 
energy of twin boundaries results in a number of superior properties over conventional grain boundaries. For example, 
it has been found that the twin boundaries enhance the strength without loss of ductility \cite{strength-1,strength-2,
strength-3}, improve fracture toughness and crack resistance \cite{crack-1,crack-2,crack-3}, and increase corrosion 
resistance and strain rate sensitivity \cite{srate}. Moreover, the twin boundaries possess high thermal and mechanical 
stability \cite{T-stability-1,T-stability-2} and high electrical conductivity \cite{E-conduct}. The superior mechanical 
properties of twinned nanopillars have been attributed to unique deformation mechanisms operating in the presence of twin
boundaries \cite{modes}. In view of this, the materials containing high density of twin boundaries attract huge interest 
among materials scientists and engineers.

Several experimental and molecular dynamics (MD) simulation studies have been performed to understand the influence of 
twin boundaries on the strength and deformation behaviour in FCC nanopillars/nanowires \cite{Cao,Zhang,Hammami,Wei,Sainath}. 
Using MD simulations, Cao et al. \cite{Cao} have shown that in FCC nanopillars, the twin boundaries serve as the strong 
obstacles for dislocations motion. As a result, the decrease in twin boundary spacing increases the yield strength in 
orthogonally twinned Cu nanopillars. Apart from obstacle to dislocation motion, the twin boundaries also serve as dislocation 
source once they lose their coherency at large plastic deformation \cite{Cao}. This nature of twin boundaries as a dislocation 
source and also as a glide plane contributes to the improvement in tensile ductility. Zhang and Huang \cite{Zhang} have shown 
that the twin boundaries do not always strengthen the nanowires. It has been demonstrated that the presence of twin boundaries 
in square cross-section nanopillars leads to strengthening effect, while in nanopillars with circular cross-sections, softening
is observed \cite{Zhang}. Further, the strengthening in twinned FCC nanopillars also depends on the size and aspect ratio 
\cite{Hammami}. In addition to orthogonally twinned FCC nanopillars, the increase in yield stress is also observed in slanted 
and vertically twinned FCC nanopillars \cite{Wei,Sainath}. Using molecular dynamics simulations and in situ experiments, the 
deformation mechanisms responsible for superior properties and the associated dislocation-twin boundary interactions have been 
characterised in FCC nanopillars.

Most of the studies reported in the literature have been performed on twinned FCC nanopillars and little attention has been 
paid to characterise the mechanical behaviour of twinned BCC nanowires/nanopillars. It is well known that the twin boundaries 
in FCC system coincide with \{111\} planes, while in BCC systems they coincide with \{112\} planes \cite{Hirth}. The FCC(111) 
plane has a closed packed structure and the twin boundary on this plane results from the stacking disorder such as ABCACBA. 
On the other hand, the BCC(112) plane is not a closed packed structure and the twin boundary consists of stacking disorder 
on \{112\} planes with six layers ABCDEFAFEDCBA. Because of the different atomic densities and stacking sequence, the twin 
boundary in BCC system possesses higher energy than the corresponding twin boundary in FCC system \cite{Japan}. Moreover, 
the mirror symmetry across the boundary plane is preserved in FCC systems. Whereas in BCC systems, the twin boundary can have 
either reflection structure, where mirror symmetry is preserved or it can have displaced structure, where the upper grain is 
displaced with respect to the lower grain by a vector 1/12$<$111$>$ \cite{Yamaguchi}. The displaced boundary no longer possesses 
the mirror symmetry. In view of different interface energies and twin boundary structures, the effect of twin boundaries in 
BCC systems may be different than that in FCC systems. For the first time, an attempt has been made in the present investigation 
to study the influence of twin boundaries on the deformation behaviour under tensile and compressive loadings in BCC nanopillars 
using atomistic simulations.

An examination of the deformation behaviour of twinned nanopillars also offers valuable insights into twin-twin and 
dislocation-twin interactions. In FCC metals, the dislocation-twin boundary and twin-twin interactions are well understood, 
and can be described by the notation of double Thompson tetrahedron  \cite{Sehitoglu,Zhu-acta}. Depending on dislocation 
character and applied stress, various reactions such as dislocation transmission across the twin boundary, sessile stair-rod 
formation, dislocation incorporation into twin boundaries leading to twin growth or detwinning and dislocation multiplication 
have been observed in FCC systems \cite{Sainath,Sehitoglu,Zhu-acta}. The absence of such simplified notation and non-planer 
core of screw dislocations along with the presence of twinning-antitwinning sense on \{112\} planes make it difficult to 
understand the dislocations-twin boundary interactions in BCC metals. There are only a couple of studies pertaining to the 
dislocation-twin and twin-twin interactions in BCC metals \cite{Mrovec,Ojha,JAP-Ta}. In view of the above observations, the 
present paper is aimed at understanding the role of twin boundaries and the mechanisms responsible for strengthening or 
softening behaviour in BCC Fe nanopillars. It is also aimed at characterising the twin-twin and dislocation-twin interactions 
observed during the deformation. The twin boundaries have been introduced perpendicular to the loading direction and number 
of twin boundaries varied from one to five with corresponding twin boundary spacing in the range 2.8 - 8.5 nm. The stress-strain 
behaviour and the variations in yield strength and deformation mechanism with respect to twin boundary spacing and loading mode 
have been discussed.

\section{MD Simulation details}

Molecular dynamics simulations have been carried out in large-scale atomic/molecular massively parallel simulator 
package \cite{Plimpton-1995} and the visualisation of atomic structure is accomplished using AtomEye \cite{J-Li-2003} 
package with centro-symmetry parameter \cite{CSP}. The embedded atom method (EAM) potential for BCC Fe given by 
Mendelev and co-workers \cite{Mendelev-2003} has been chosen to describe the interaction between Fe atoms. This 
potential is widely used to study the deformation behaviour of BCC Fe \cite{Kotrechko,Healy-2015,Sainath-CMS15,
Sainath-MSEA,Sainath-CMS16}. In order to create twinned nanopillars, the following procedure was adopted. Initially, 
the single crystal BCC Fe nanopillars of square cross section width (d) = 8.5 nm and consisting of about 110,000 atoms 
oriented in $<$112$>$ axial direction with \{110\} and \{111\} as side surfaces was constructed. The nanopillar length 
(l) was twice the cross section width (d). Following this, the twin boundaries were introduced by rotating one part of 
the crystal with respect to other by $180^o$ around $<$112$>$ axis. Following the rotation, the twin boundary forms at 
their interface on \{112\} plane. The formed twin boundary is equivalent to a twist boundary lying on \{112\} plane 
with a twist angle of $180^o$. Similar procedure was followed to create more number of twin boundaries. The model 
system was equilibrated to a temperature of 10 K in NVT ensemble. In all the nanopillars, no periodic boundary 
conditions were used in any direction. Following the relaxation, the twin boundary having displaced structure was 
observed  \cite{Yamaguchi,Sainath-CMS15,Sainath-CMS16,Scripta-16}. The nanopillars containing one, two, three and five 
twin boundaries resulted in the twin boundary spacings of 8.5, 5.7, 4.2 and 2.8 nm, respectively. The BCC Fe nanopillars 
with different twin boundary spacings considered in this study along with perfect nanopillar are shown in Figure 
\ref{Initial}. Since the present study focuses on the effect of twin boundary spacing on deformation behaviour and 
therefore, the effects associated with nanopillar size, aspect ratio and temperature have not been considered. Upon 
completion of equilibrium process, the deformation under tensile and compressive loadings was carried out in a displacement 
controlled mode at a constant strain rate of $1 \times 10^8$ s$^{-1} $  by imposing displacements to atoms along the 
nanopillar length that varied linearly from zero at the bottom to a maximum value at the top layer. The average stress 
is calculated from the Virial expression \cite{Virial}.

 \begin{figure}[h]
\centering
 \includegraphics[width= 14cm]{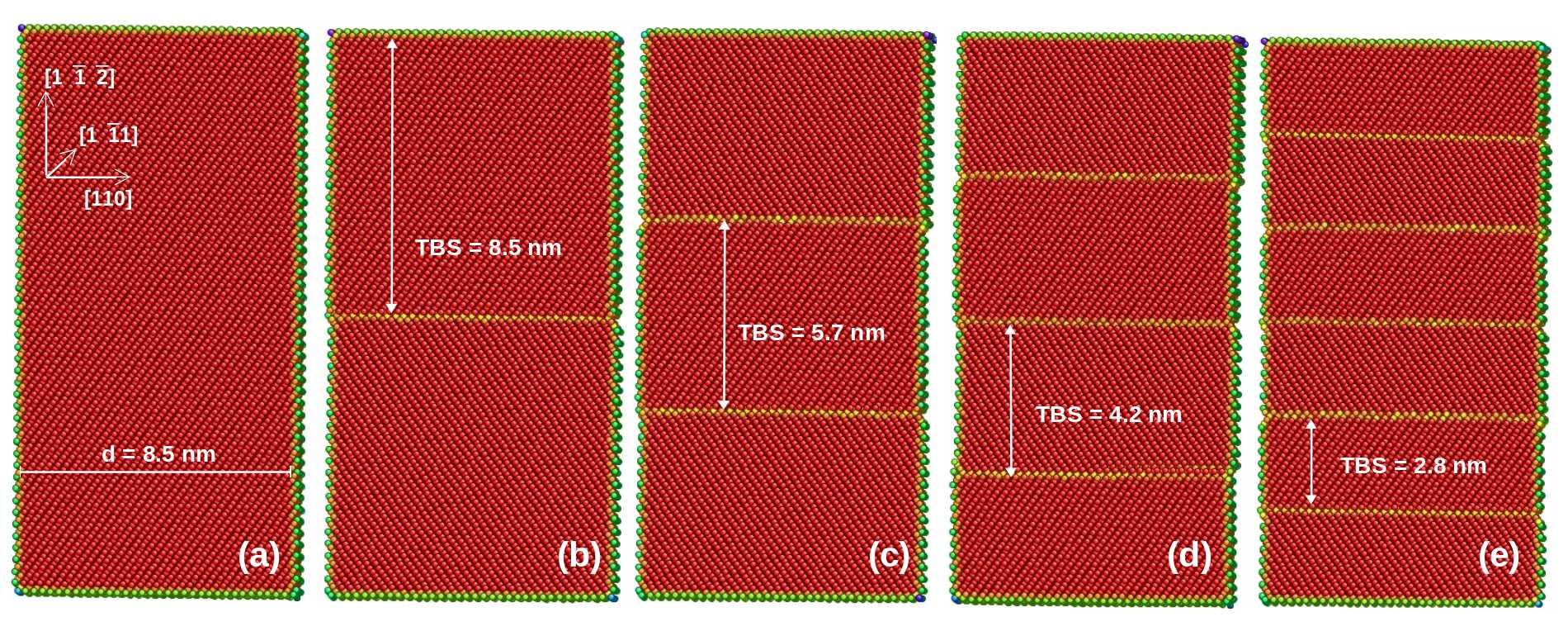}
 \caption {\footnotesize The initial configuration of (a) perfect BCC Fe nanopillar and the nanopillar consisting of 
 (b) one (c) two (d) three and (e) five twin boundaries. The corresponding twin boundary spacings (TBS) of 8.5, 5.7, 4.2 
 and 2.8 nm are shown. The atoms are coloured according to the centro-symmetry parameter \cite{CSP}.}
 \label{Initial}
 \end{figure}
 
The Burgers vectors of full dislocations were determined using Dislocation extraction algorithm (DXA) developed by 
Stukowski \cite{DXA-2010} as implemented in OVITO \cite{OVITO}. However, the identification of Burgers vector of 
partial/twinning dislocations gliding on the twin boundary in BCC system is challenging. This is mainly because the 
Burgers circuit of twinning dislocation passes through the twin boundary at which the crystal orientation changes.
Therefore, the reference frame for the Burgers vector analysis is no longer the perfect lattice, but is a bicrystal 
containing a perfect twin boundary \cite{DXA-2010}. Due to this difficulty, the DXA or OVITO packages failed to detect 
the Burgers vectors of partial dislocations and the type of dislocation intersection. We have assigned the Burgers 
vector of partial/twinning dislocations in BCC Fe based on the experimental observations of Paxton \cite{Paxton}. 
Generally in BCC systems, the 1/6$<$111$>$ type of partial dislocations are responsible for twinning mechanism 
\cite{Hirth}. Therefore, we have taken 1/6$<$111$>$ as the Burgers vector of partial dislocations gliding along the 
twin boundary. Finally, the type of twin-twin intersection is identified based on the common line of intersection 
\cite{Mahajan}. If m and n are the plane normals of the two interacting twins, the intersection type is the vector 
corresponding to their common intersection line, and is given by the cross product of m and n \cite{Ojha}.

\section{Results}

\subsection{Stress-strain behaviour}

The stress-strain behaviour of BCC Fe nanopillars under tensile loading containing one, two, three and five twin 
boundaries along with the perfect nanopillar is shown in Figure \ref{stress-strain-tension}. All the nanopillars 
exhibited linear elastic deformation at small strains followed by nonlinearity at higher strains, i.e. $\varepsilon 
> 0.05$. The modulus evaluated from the linear elastic regime displayed insignificant variations and this indicated 
that the elastic modulus is not influenced by the presence of twin boundaries. Following elastic deformation, the 
large and abrupt drop in flow stress signifying the occurrence of yielding in the perfect and twinned nanopillars 
is seen in Figure \ref{stress-strain-tension}. In BCC Fe, introduction of twin boundaries resulted in the significant
reduction in yield stress compared to that in perfect nanopillar. Similarly, a decrease in the strain to yielding 
has also been observed in the twinned nanopillars. The yield stress exhibited only marginal variation with respect 
to twin boundary spacing as shown in Figure \ref{TBS-Tension}. For comparison, the yield stress of perfect nanopillar 
is shown as horizontal line in Figure \ref{TBS-Tension}. Following yielding, the perfect nanopillar exhibited nearly 
a constant and low flow stress during plastic deformation up to large strains. Contrary to this, the nanopillars 
containing one and two twin boundaries displayed a gradual decrease in the flow stress with increase in plastic 
deformation. A rapid decrease in flow stress with increase in plastic strain was observed for nanopillars having 
three and five twin boundaries. Further, the nanopillars having three and five twin boundaries shows higher flow 
stress with significant fluctuations at low plastic strains. A general decrease in strain to failure has been obtained 
in the twinned nanopillars compared to that in the perfect nanopillar.

\begin{figure}[h]
\centering
 \includegraphics[width= 8.5cm]{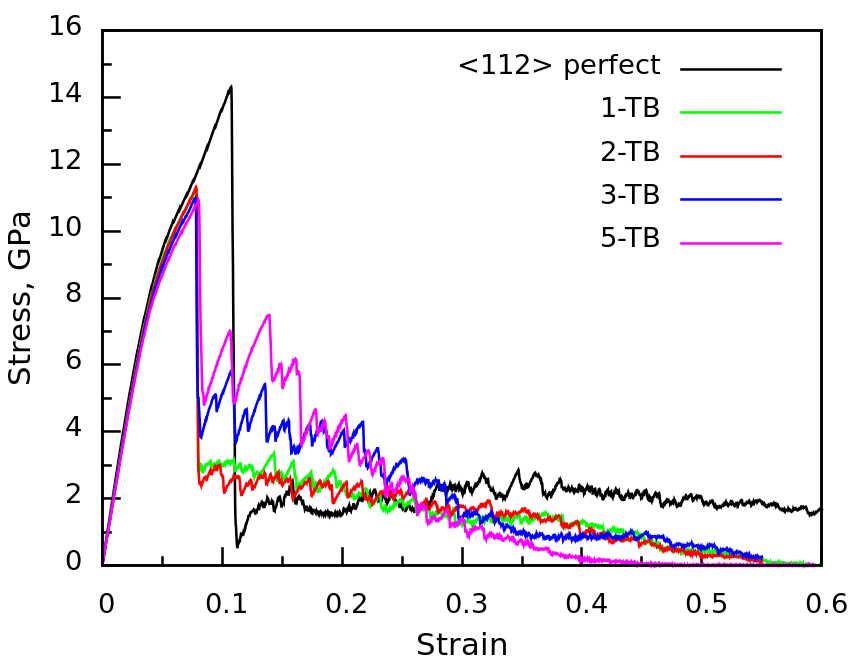}
 \caption {\footnotesize The stress-strain behaviour of BCC Fe nanopillars containing one, two, three and five 
 twin boundaries (TBs) along with stress-strain behaviour of perfect nanopillar under tensile loading.}
 \label{stress-strain-tension}
 \end{figure}
 
  \begin{figure}[h]
\centering
 \includegraphics[width= 8.5cm]{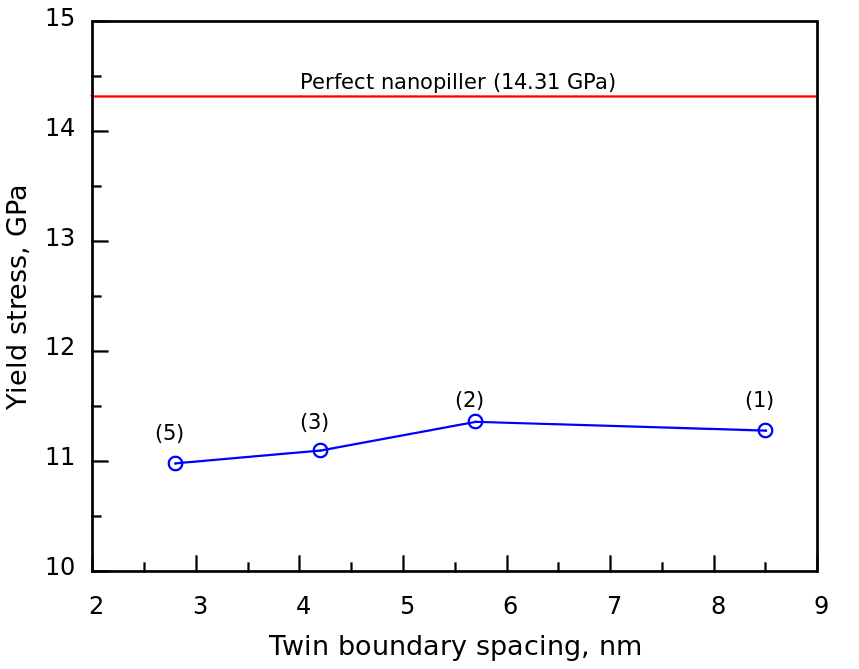}
 \caption {\footnotesize Variation in yield stress as a function of twin boundary spacing in BCC Fe nanopillars under 
 tensile loading. The yield stress of perfect nanopillar is indicated by red line and the number of twin boundaries is 
 shown in small brackets.}
 \label{TBS-Tension}
 \end{figure}

Figure \ref{stress-strain-compress} shows the stress-strain behaviour of perfect and twinned BCC Fe nanopillars under 
compressive loading. It can be seen that the defect free nanopillar exhibits perfect linear elastic deformation, while 
the twinned nanopillars display linear elastic deformation at small strains followed by non-linear elastic deformation 
at high strains. In addition to this, the perfect nanopillar exhibited higher elastic modulus than those observed for 
twinned nanopillars. Following elastic deformation, the nanopillars displayed yielding characterised by an abrupt drop 
in flow stress. However, the drop in flow stress during yielding under compressive loading (Figure \ref{stress-strain-compress}) 
has been significantly lower than those under tensile loading (Figure \ref{stress-strain-tension}). Under compressive 
deformation, decrease in yield stress in the presence of a single twin boundary followed by an increase in yield stress 
with increase in the number of twin boundaries has been observed. Finally, the yield stress attains a value closer (with
three twin boundaries) or marginally higher (with five twin boundaries) than that in the perfect nanopillar (Figure 
\ref{TBS-Compression}). In general, under compressive loading, all the nanopillars displayed large oscillations and 
gradual decrease in flow stress with progressive plastic deformation (Figure \ref{stress-strain-compress}).
 
 \begin{figure}
\centering
 \includegraphics[width= 8.5cm]{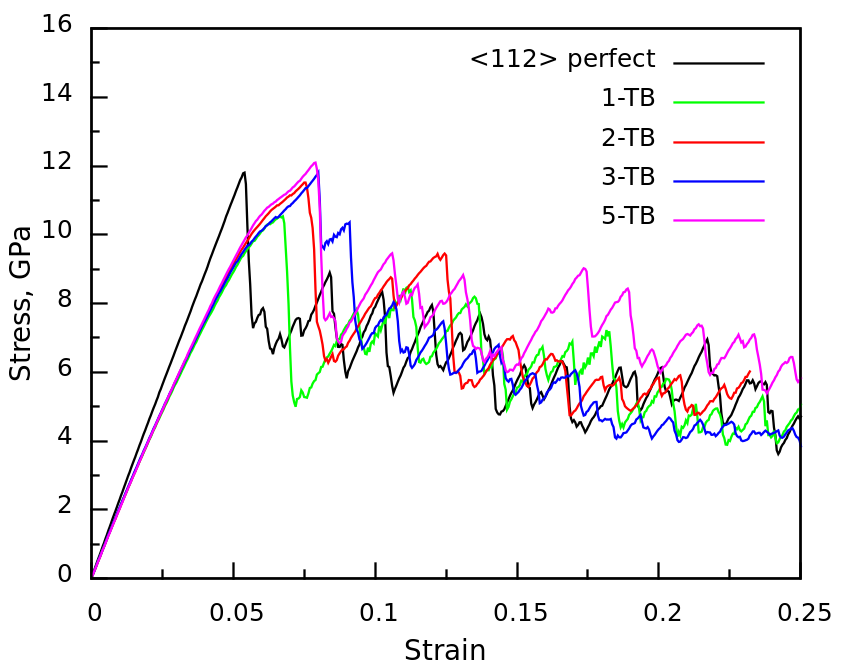}
 \caption {\footnotesize The stress-strain behaviour of BCC Fe nanopillars containing one, two, three and five twin 
 boundaries (TBs) along with stress-strain behaviour of perfect nanopillar under compressive loading.}
 \label{stress-strain-compress}
 \end{figure}
 
 \begin{figure}
\centering
 \includegraphics[width= 8.5cm]{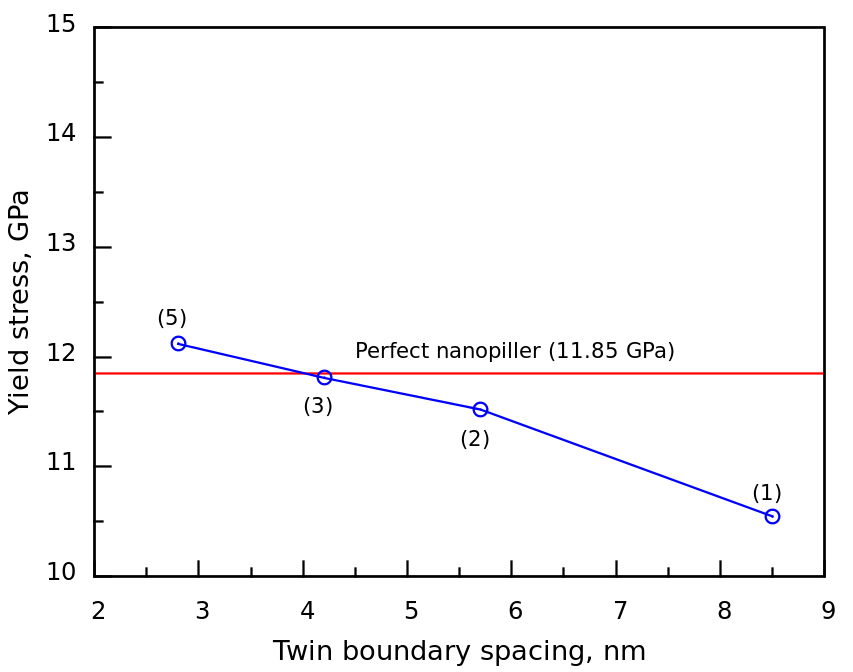}
 \caption {\footnotesize Variation in yield stress as a function of twin boundary spacing in BCC Fe nanopillars under 
 compressive loading. The yield stress of perfect nanopillar is indicated by red line and the number of twin boundaries 
 is shown in small brackets.}
 \label{TBS-Compression}
 \end{figure}
  
\subsection{Deformation behaviour under tensile loading}

The evolution of atomic configurations at various stages of deformation during the tensile loading of the $<$112$>$ 
perfect and twinned BCC Fe nanopillars has been analysed using centro-symmetry parameter \cite{CSP}. It has been observed that 
the deformation is dominated mainly by the nucleation of twin embryo from the corner during yielding followed by twin 
growth during plastic deformation. In the detailed investigation performed on $<$112$>$ perfect BCC Fe nanowires, it 
has been demonstrated that the yielding and subsequent plastic deformation occur by twinning mechanism along with 
minor dislocation slip activity \cite{Sainath-CMS16}. Due to deformation twinning, the original $<$112$>$ nanowire 
with \{111\} and \{110\} surfaces transforms to $<$100$>$ nanowire with \{100\} lateral surfaces \cite{Sainath-CMS16}. 
Similar to perfect nanopillar, the deformation in twinned nanopillars is dominated by the twinning and associated 
partial dislocation mechanism. Figure \ref{deform-tension} shows the deformation behaviour under the tensile loading 
of BCC Fe nanopillar containing a single twin boundary. The nanopillar yields by the nucleation of a twin embryo from 
the intersection of surface and the existing twin boundary (Figure \ref{deform-tension}(a)). Similar yielding behaviour 
was observed in nanopillars containing higher number of twin boundaries. With small increase in strain, the nucleation 
and glide of 1/6 $<$111$>$ partial dislocation along the initial twin boundary can be seen in Figure \ref{deform-tension}(b). 
The glide of partial dislocation leads to the migration of initial twin boundary and thereby changes the twin boundary 
spacing (Figure \ref{deform-tension}(c) and (d)). The detailed mechanism of twin boundary formation and the migration 
of initial twin boundary are shown in Figures \ref{Twin-growth} and \ref{TBM} respectively. Initially the 1/6$<$111$>$ 
partial dislocation nucleates on \{112\} plane from the corner of the nanowire with a stacking fault behind (Figure 
\ref{Twin-growth}(a)). Upon increasing deformation, this partial dislocation glides further in $<$111$>$ direction 
(Figure \ref{Twin-growth}(b)) and an additional 1/6$<$111$>$ partial dislocations nucleates from the intersection 
of the surface and stacking fault as shown in Figure \ref{Twin-growth}(b) and (c). When twin front reaches the opposite 
surface, the twin embryo eventually becomes full twin enclosed by two \{112\} twin boundaries (Figure \ref{Twin-growth}(d)). 
Following this, the twin grows along the nanowire axis by the successive nucleation and glide of 1/6$<$111$>$ partial 
dislocations on adjacent \{112\} planes. Figure \ref{TBM} shows the migration of initial twin boundary due to glide of 
1/6$<$111$>$ partial dislocations, which leads to change in twin boundary spacing. Each nucleation, glide and annihilation 
of 1/6$<$111$>$ partial dislocation displaces the twin boundary by one layer. As a result of repeated nucleation and glide, 
the twin boundary migrates marginally along the nanopillar axis (Figure \ref{deform-tension}(d)). Although, the overall 
deformation is dominated by twinning mechanism (Figure \ref{deform-tension}(d)), minor activity of full dislocation slip 
is also observed in the neighbouring grain (Figure \ref{deform-tension}(c)). However, this full dislocation slip contributes 
negligibly to the overall strain. Similar to nanopillar having a single twin boundary, deformation by twinning occurs in 
nanopillars containing higher number of twin boundaries (Figure \ref{All-TB-Tension}). However, few important differences 
were noticed in nanopillars containing two, three and five twin boundaries. It can be seen that in nanopillar containing 
two twin boundaries, the twinning occurs only in two grains and no activity of slip is observed in any grain 
(Figure \ref{All-TB-Tension}(a)). 
Moreover, the operative twin systems in these two grains are symmetrical with respect to twin boundary. In contrast, in 
nanopillars containing three and five twin boundaries, considerable activity of full dislocations is also observed 
(Figure \ref{All-TB-Tension}(b) and (c)). This full dislocation activity is aided by twin-twin interactions. Furthermore, the 
change in twin boundary spacing increases with increase in the number of twin boundaries. In addition, in nanopillars 
containing five twin boundaries, the twin boundary spacing has become uneven due to continuous glide of parallel twinning 
partials (Figure \ref{All-TB-Tension}(c)). Upon increasing strain, the deformation is mainly concentrated at the twin 
boundaries in all the nanopillars, leading to the occurrence of necking close to one of the twin boundaries (Figure 
\ref{All-TB-Tension}).

 \begin{figure}
\centering
 \includegraphics[width= 12cm]{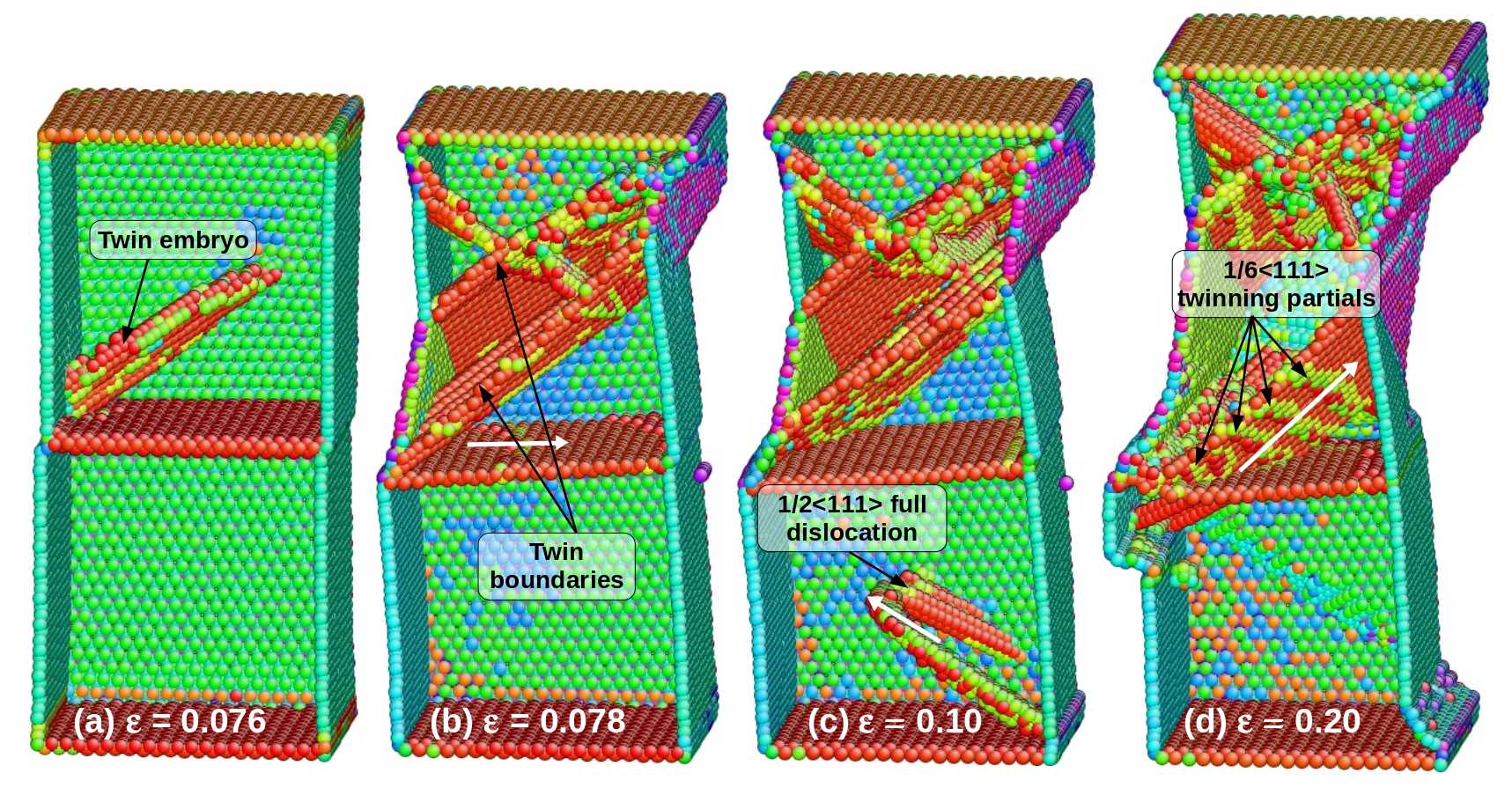}
 \caption {\footnotesize The deformation behaviour of BCC Fe nanopillar containing single twin boundary under tensile 
 loading. The atoms are coloured according to the centro-symmetry parameter \cite{CSP}. The perfect BCC atoms and the 
 front surface are removed for clarity.}
 \label{deform-tension}
 \end{figure}
 
\begin{figure}
\centering
 \includegraphics[width= 10cm]{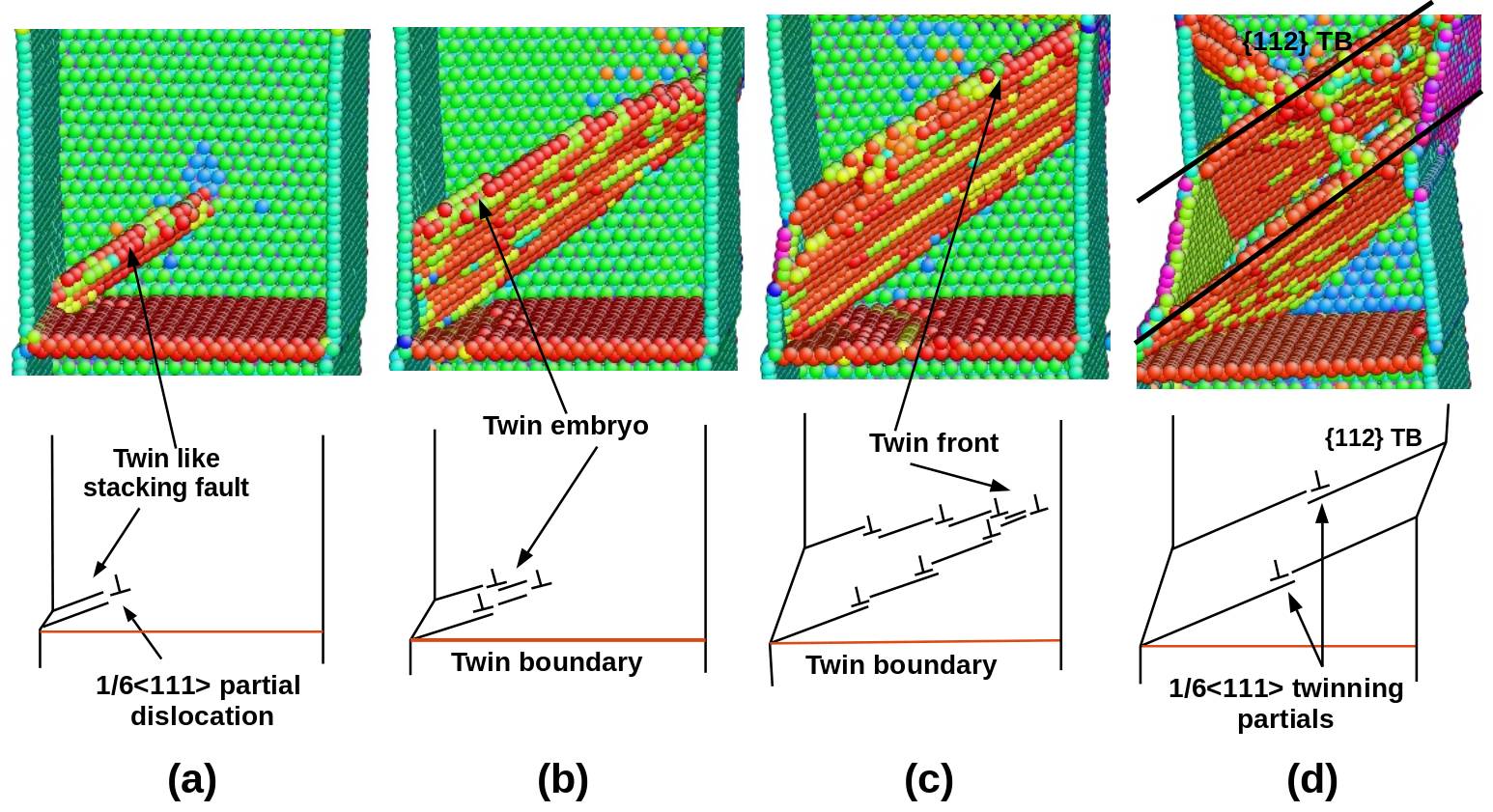}
 \caption {\footnotesize The detail process of twin embryo nucleation to twin boundary formation. The 2-D view of twin 
 nucleation and growth is shown schematically in lower figures. The atoms are coloured according to the centro-symmetry 
 parameter \cite{CSP}.}
 \label{Twin-growth}
 \end{figure}

\begin{figure}[h]
\centering
 \includegraphics[width= 9.5cm]{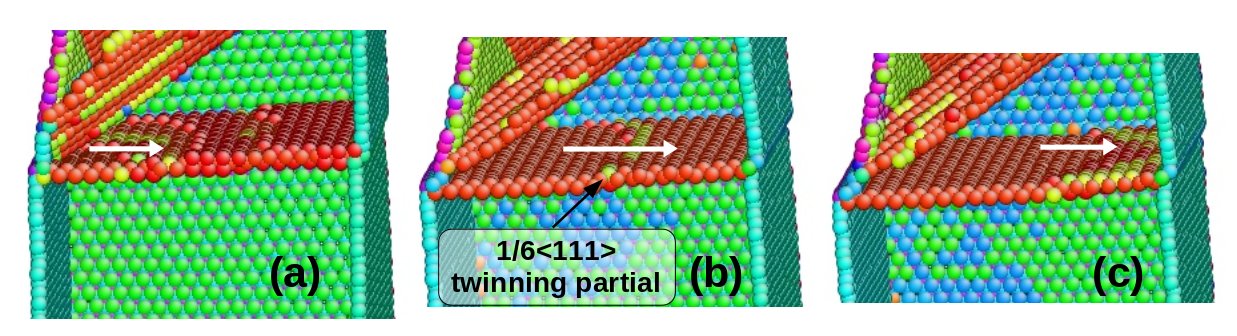}
 \caption {\footnotesize Typical glide of partial dislocations along the existing twin boundary. The continuous nucleation 
 and glide of partial dislocations migrating the initial twin boundary is shown. The atoms are coloured according to the 
 centro-symmetry parameter \cite{CSP}. The perfect BCC atoms and the front surface are removed for clarity.}
 \label{TBM}
 \end{figure}

 \begin{figure}[h]
\centering
 \includegraphics[width= 14cm]{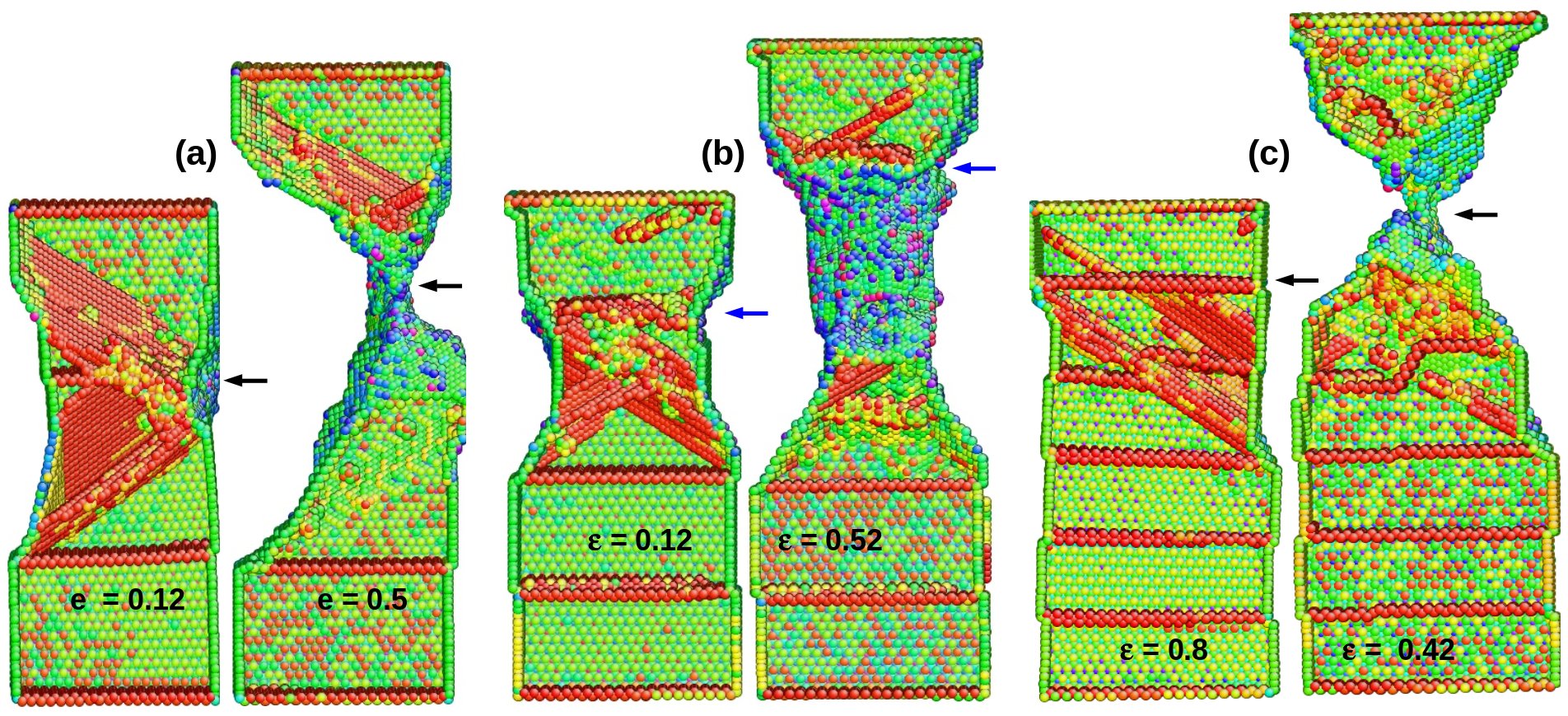}
 \caption {\footnotesize Deformation behaviour of BCC Fe nanopillars containing (a) two, (b) three and (c) five twin 
 boundaries. The atoms are coloured according to the centro-symmetry parameter \cite{CSP}. The perfect BCC atoms and 
 the front surface are removed for clarity.}
 \label{All-TB-Tension}
 \end{figure}
 
As the deformation under tensile loading of twinned nanopillars is dominated by twinning mechanism, it can offers an 
insights into twin-twin interactions. Figure \ref{cross-slip} shows the twin-twin interactions under the tensile loading 
of twinned nanopillars observed in present study. Initially, the twin embryo consisting of many partial dislocations 
nucleates from the corner of the nanopillar 
and propagates towards the initial twin boundary (Figure \ref{cross-slip}(a)). The nucleated twin on [112] 
plane interacts with initial twin boundary on $[1\bar1\bar2]$ plane and produces $<$012$>$ twin-twin 
intersection (Figure \ref{cross-slip}(b)). In all the nanopillars undergoing twinning, $<$012$>$ type twin-twin 
intersections have been observed. Due to this twin-twin interaction, the formation of the twinning partials 
gliding along the twin boundary can be seen in Figure \ref{cross-slip}(b) and (c). The other possibility of 
twin-twin interactions arise from the pile-ups and combination of twinning partial dislocations (Figure 
\ref{cross-slip}(d)). During deformation, the gliding twinning partials along the nucleated twin boundary 
pile-up against the initial twin boundary (Figure \ref{cross-slip}(d)). Following the pile-ups, three of 
twinning partials combine and form a full dislocation (Figure \ref{cross-slip}(e)). Upon increasing strain, 
the lattice near the twin-twin intersection is highly distorted (Figure \ref{cross-slip}(f)). The full 
dislocations thus nucleated, glide on the plane symmetrical with respect to the plane of nucleated deformation 
twin.

\begin{figure}
\centering
\includegraphics[width= 8.5cm]{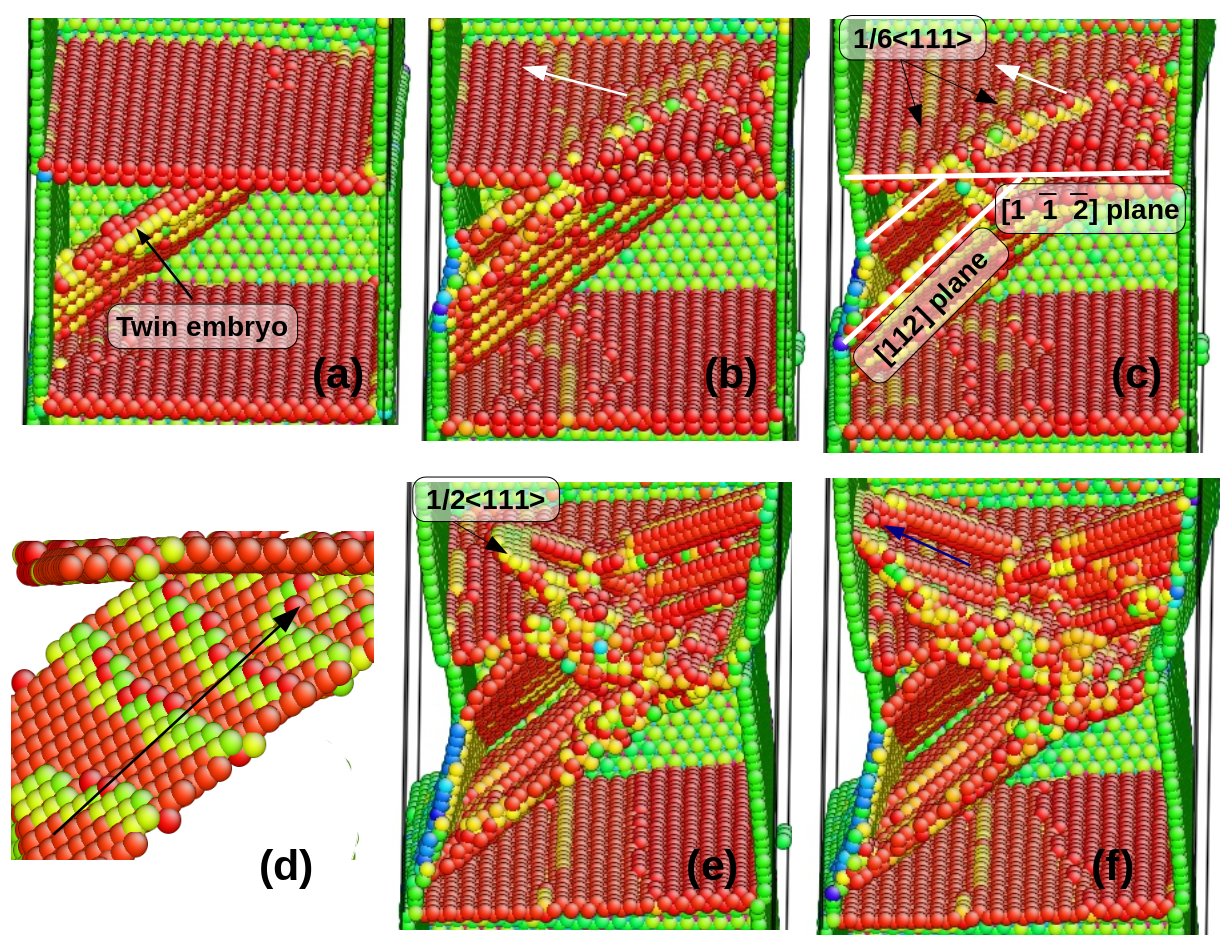}
\caption {\footnotesize Twin-twin interaction under the tensile loading of twinned BCC Fe nanopillars. The nucleated 
twin on [112] plane interacts with the initial twin boundary on  $[1\bar1\bar2]$ plane and produces a $<$012$>$ type 
twin-twin intersection. $<$012$>$ twin-twin intersection is obtained as the cross product of (112) and (1-1-2). The 
full dislocation emission from the twin-twin interaction can be seen in (c) and (d). The atoms are coloured according 
to the centro-symmetry parameter \cite{CSP}. The perfect BCC atoms and the front surface are removed for clarity.}
\label{cross-slip}
\end{figure}

\subsection{Deformation behaviour under compression}
 
 \begin{figure}[h]
\centering
 \includegraphics[width= 7cm]{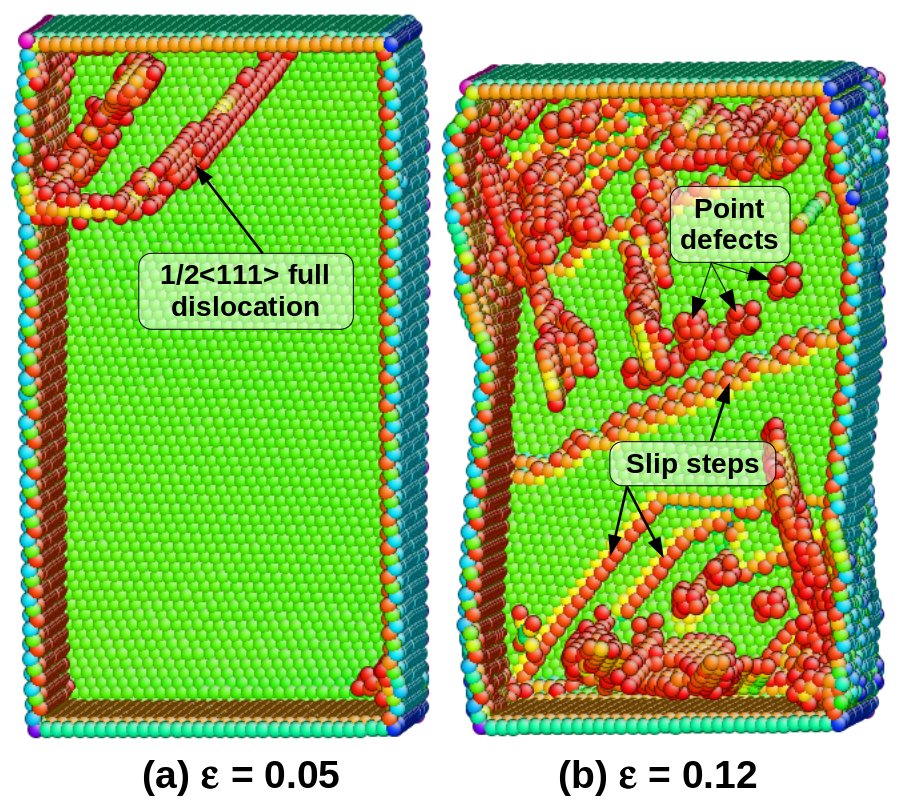}
 \caption {\footnotesize Deformation behaviour of perfect $<$112$>$ BCC Fe nanopillar under compressive loading. The atoms 
 are coloured according to the centro-symmetry parameter \cite{CSP}. The perfect BCC atoms and the front surface are removed 
 for clarity.}
 \label{perfect-compress}
 \end{figure}

The deformation behaviour under the compressive loading of perfect and twinned BCC Fe nanopillars is 
dominated mainly by the slip of full dislocations. Figure \ref{perfect-compress} shows the deformation behaviour 
of perfect BCC Fe nanopillar oriented in $<$112$>$ axis under compressive loading. It can be seen that 
the yielding occurs by the nucleation of 1/2$<$111$>$ full dislocations from the corner of the nanopillar 
(Figure \ref{perfect-compress}(a)). Following yielding, the plastic deformation is entirely dominated 
by the slip of full dislocations. Due to glide of 1/2$<$111$>$ screw dislocations, the straight and curved 
slip steps were observed on surface of the nanopillar (Figure \ref{perfect-compress}(b)). Furthermore, the 
formation of large number of point defects can be seen in Figure \ref{perfect-compress}(b). The non-conservative 
motion of dislocations leads to the creation of point defects such as vacancies and/or interstitials 
\cite{Sainath-MSEA}. The typical atomic configurations representing plastic deformation under compressive 
loading of twinned nanopillars containing two twin boundaries are shown in Figure \ref{deform-compress}. 
The onset of yielding is characterised by the nucleation of dislocation loop originating from the corner 
of the nanopillar (Figure \ref{deform-compress}(a)). The DXA analysis \cite{DXA-2010} indicated that the 
dislocation loop has a Burgers vector 1/2$<$111$>$ representing full dislocations in BCC system. Following 
nucleation, the dislocation loop expands in diameter and the part of the loop is annihilated at the free 
surface, while the remaining part is blocked at twin boundary (Figure \ref{deform-compress}(b)). This 
indicates that the twin boundaries in BCC Fe nanopillars are effective barriers for dislocation motion. 
With the increase in stress, the blocked dislocation penetrates the twin boundary and comes out as a loop 
in the next grain (Figure \ref{deform-compress}(c)). During this process of nucleation and propagation,
the accumulation of large of number of straight screw dislocations can be seen at higher strains in all 
the nanopillars (Figure \ref{All-TB}(a)–(d)). The accumulation process of straight screw dislocations in 
BCC nanowires has also been demonstrated in our earlier investigation \cite{Sainath-MSEA}. It has been observed 
that the dislocation loop initially nucleated from the corner consists of edge as well as screw components. 
In view of higher mobility of edge dislocations, the edge component of mixed dislocations easily escapes 
to the surface resulting in the accumulation of screw dislocations. In nanopillars containing one and two 
twin boundaries, accumulation of long and straight screw dislocations has been observed (Figure \ref{All-TB}(a) 
and (b)), while nanopillars having three and five twin boundaries, accrual of comparatively smaller screw 
dislocations have been noticed (Figure \ref{All-TB}(c) and (d)). Furthermore, due to dislocation blockage 
by twin boundaries, the formation of hairpin-like dislocations has been observed during compressive deformation 
of twinned nanopillars with higher number of twin boundaries (Figure \ref{All-TB}(d)). Generally, the dislocations 
that glide in hairpin-like configuration are known as hairpin dislocations and this kind of dislocations have also 
been observed in twinned FCC nanocrystalline materials \cite{Hair-pin}.

 \begin{figure}[h]
\centering
 \includegraphics[width= 9.5cm]{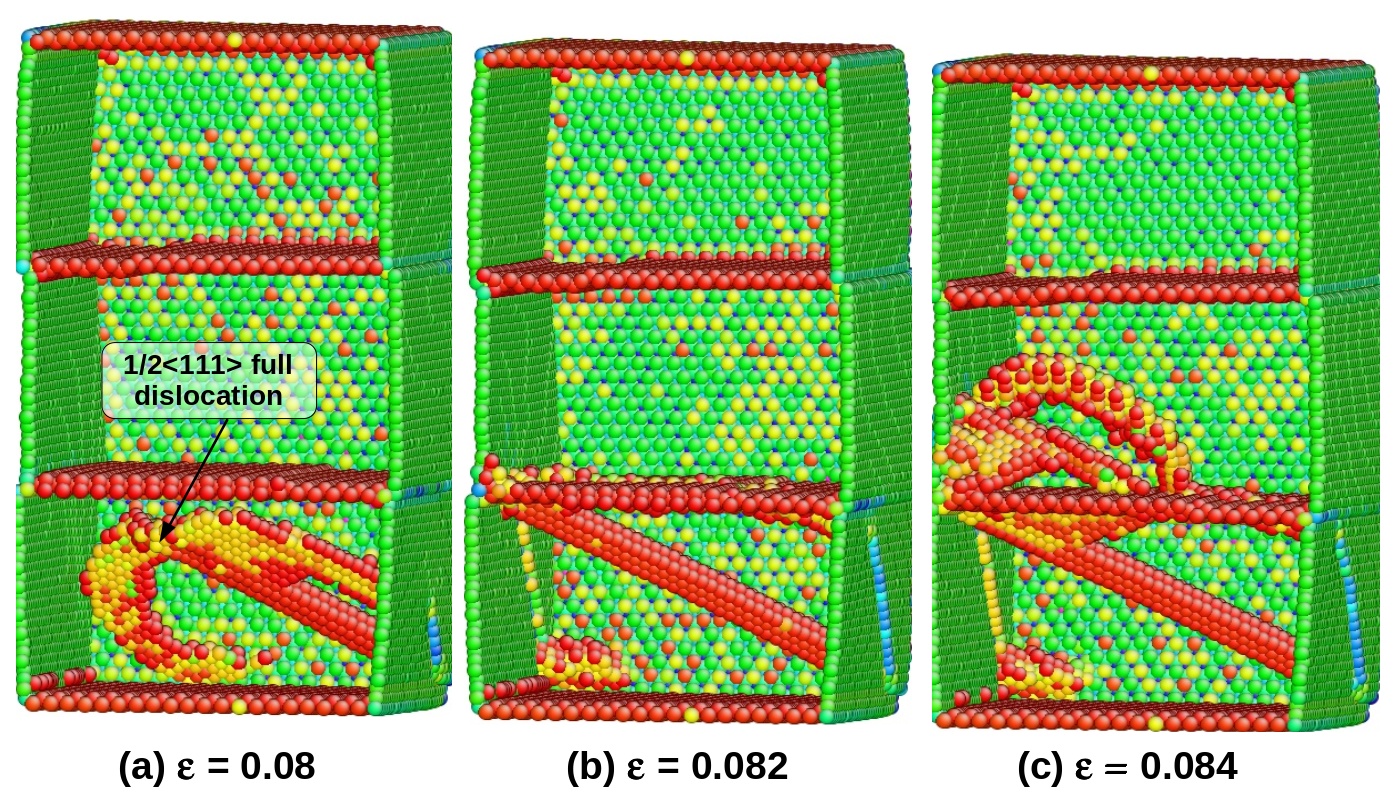}
 \caption {\footnotesize Deformation behaviour of BCC Fe nanopillar containing two twin boundaries under the 
 compressive loading. The atoms are coloured according to the centro-symmetry parameter \cite{CSP}. The perfect 
 BCC atoms and the front surface are removed for clarity.}
 \label{deform-compress}
 \end{figure}
 
  \begin{figure}[h]
\centering
 \includegraphics[width= 14cm]{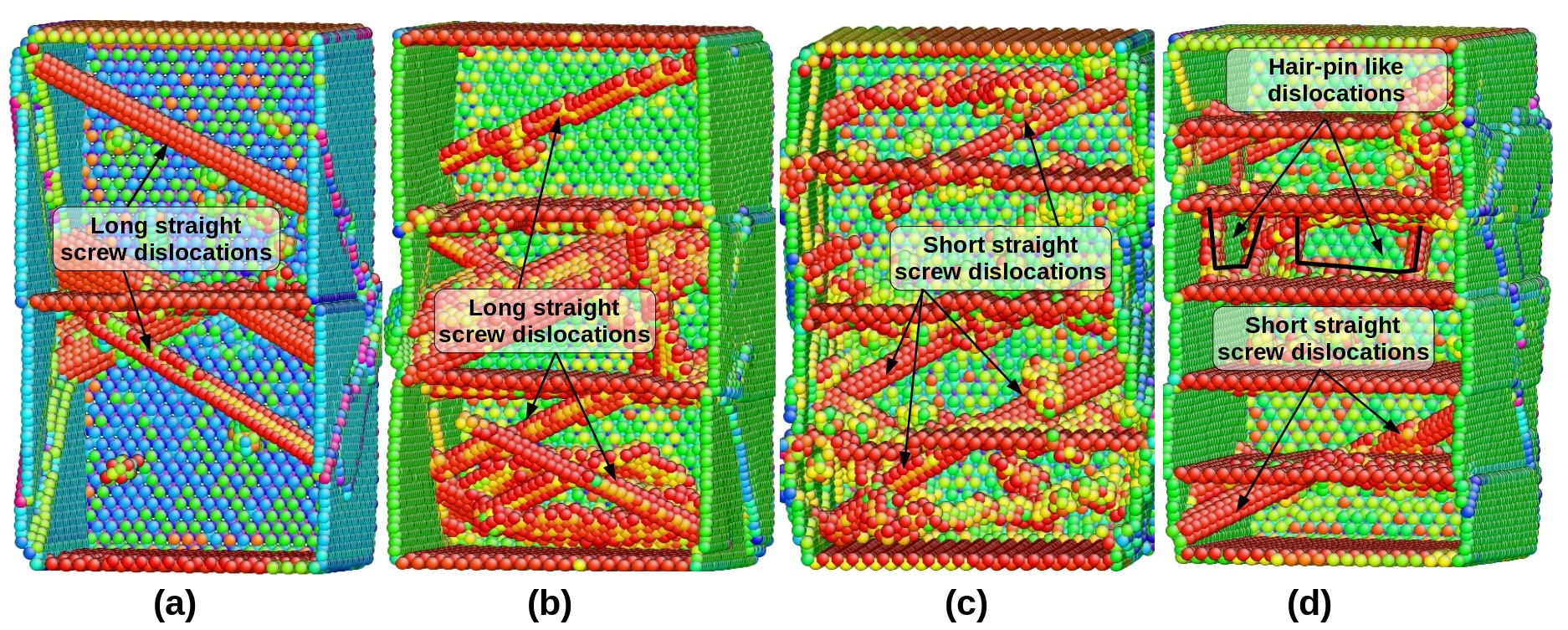}
 \caption {\footnotesize Accumulation of straight screw dislocations in the twinned BCC Fe nanopillars under the 
 compressive loading. The atoms are coloured according to the centro-symmetry parameter \cite{CSP}. The perfect 
 BCC atoms and the front surface are removed for clarity.}
 \label{All-TB}
 \end{figure}

Since the deformation of twinned nanopillars under the compressive loading is dominated by the full dislocations, 
it can offers an insights into dislocation-twin boundary interactions. In the present investigation, the dislocation-twin 
interactions have been observed for all the nanopillars under compressive loading. The direct transmission of 1/2$<$111$>$ 
full dislocation across the twin boundary without any deviation in glide plane is shown in Figure \ref{TB-inter-1}. 
Initially, 1/2$<$111$>$ full dislocation loop nucleates from the corner of nanopillar during yielding and grows with 
deformation (Figure \ref{TB-inter-1}(a)). Once the part of the dislocations reaches the twin boundary, the dislocation 
line becomes parallel to the intersection line of the glide plane and twin boundary (Figure \ref{TB-inter-1}(b)). With 
the increase in strain, this dislocation passes through the twin boundary and glides on the plane parallel to the initial 
glide plane (Figure \ref{TB-inter-1}(c)). In addition to full dislocation directly transmitting through the twin boundary 
(Figure \ref{TB-inter-1}), another operating mechanism of dislocation transmission across the twin boundary is shown in 
Figure \ref{TB-inter-2}. In this case, the dislocation comes out of the twin boundary and glides on a plane symmetrical 
to the initial glide plane (Figure \ref{TB-inter-2}(a)–(c)).

\begin{figure}[h]
\centering
\includegraphics[width= 9cm]{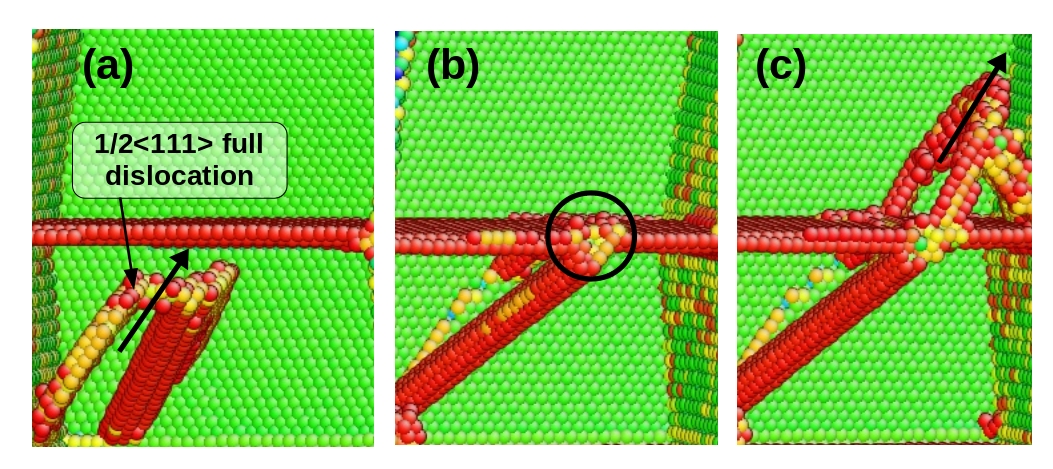}
\caption {\footnotesize Dislocation-twin boundary interaction showing direct transmission of dislocation across the 
twin boundary. The viewing direction is $<$111$>$ and the atoms are coloured according to the centro-symmetry parameter 
\cite{CSP}. The perfect BCC atoms and the front surface are removed for clarity.}
\label{TB-inter-1}
\end{figure}

\begin{figure}[h]
\centering
\includegraphics[width= 9.5cm]{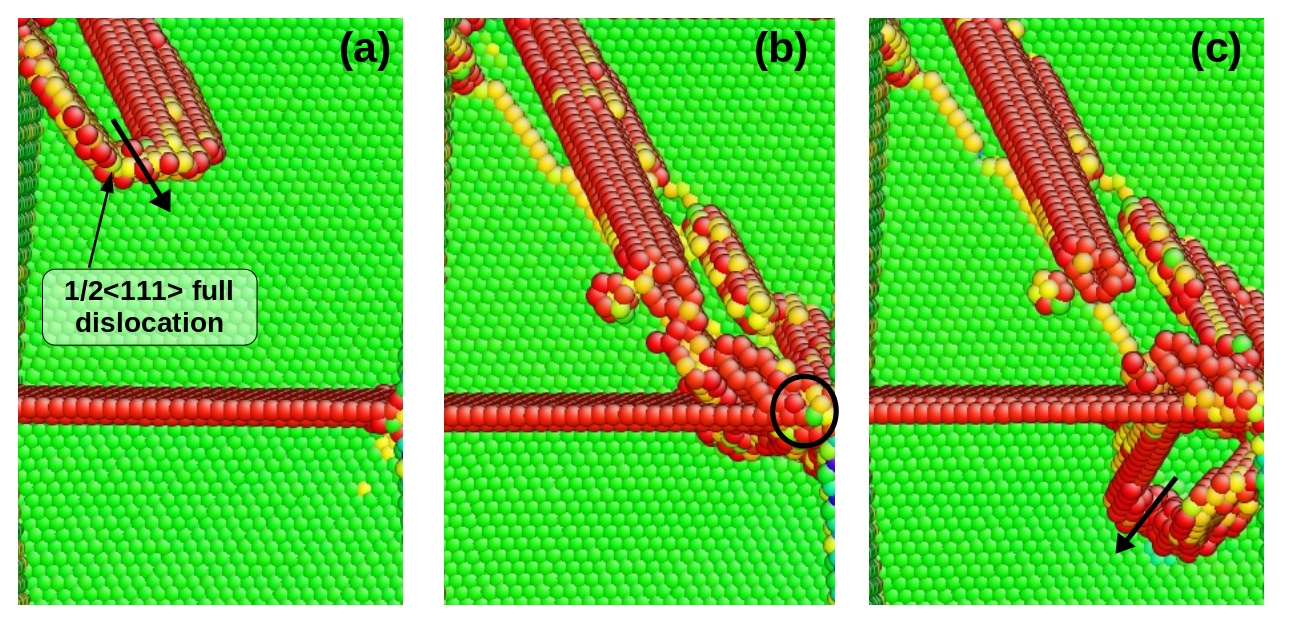}
\caption {\footnotesize Dislocation-twin boundary interaction showing the symmetrical transmission of dislocation across 
the twin boundary. The viewing direction is $<$111$>$ and the atoms are coloured according to the centro-symmetry parameter
\cite{CSP}. The perfect BCC atoms and the front surface are removed for clarity.}
\label{TB-inter-2}
\end{figure}

\subsection{Twin boundary as a dislocation source}

Under the tensile deformation of twinned BCC Fe nanopillars, it has been observed that the nucleated twin 
boundary can often be of curved nature and edge of this curved twin boundary can acts as a source for nucleation 
of full dislocations. Figure \ref{full-slip} shows pictorial view of full dislocation emission from the curved 
twin boundary. Initially, 1/6$<$111$>$ partial dislocation labelled as ‘1’ nucleates and glides along the 
curved twin boundary (Figure \ref{full-slip}(a)). The motion of this partial dislocation is prevented at the 
edge of the twin boundary, and at the same time, another partial dislocation labelled as ‘2’ nucleates and 
approaches towards the edge of twin boundary (Figure \ref{full-slip}(b)). These two partial dislocations 
combine and form a 1/3$<$111$>$ partial dislocation labelled as ‘1 + 2’ in Figure \ref{full-slip}(c). With 
a small increase in strain, one more 1/6$<$111$>$ partial dislocation labelled as ‘3’ gets nucleated, and its 
subsequent glide and combination with the existing combined 1/3$<$111$>$ partial dislocation leads to the 
formation of 1/2$<$111$>$ full dislocation as shown in Figure \ref{full-slip}(d). Thus, the nucleation of 
three successive partial dislocations and their pile up at the edge of the twin boundary leads to the formation 
of full dislocation. The 1/2 $<$111$>$ full dislocation glide on \{110\} plane and moves away from the edge of 
the twin boundary. The nucleation of dislocation from the edge of the curved twin boundary is associated with 
the reduction in the thickness of twin (Fig. \ref{full-slip}a-f). 

\begin{figure}
\centering
\includegraphics[width= 10cm]{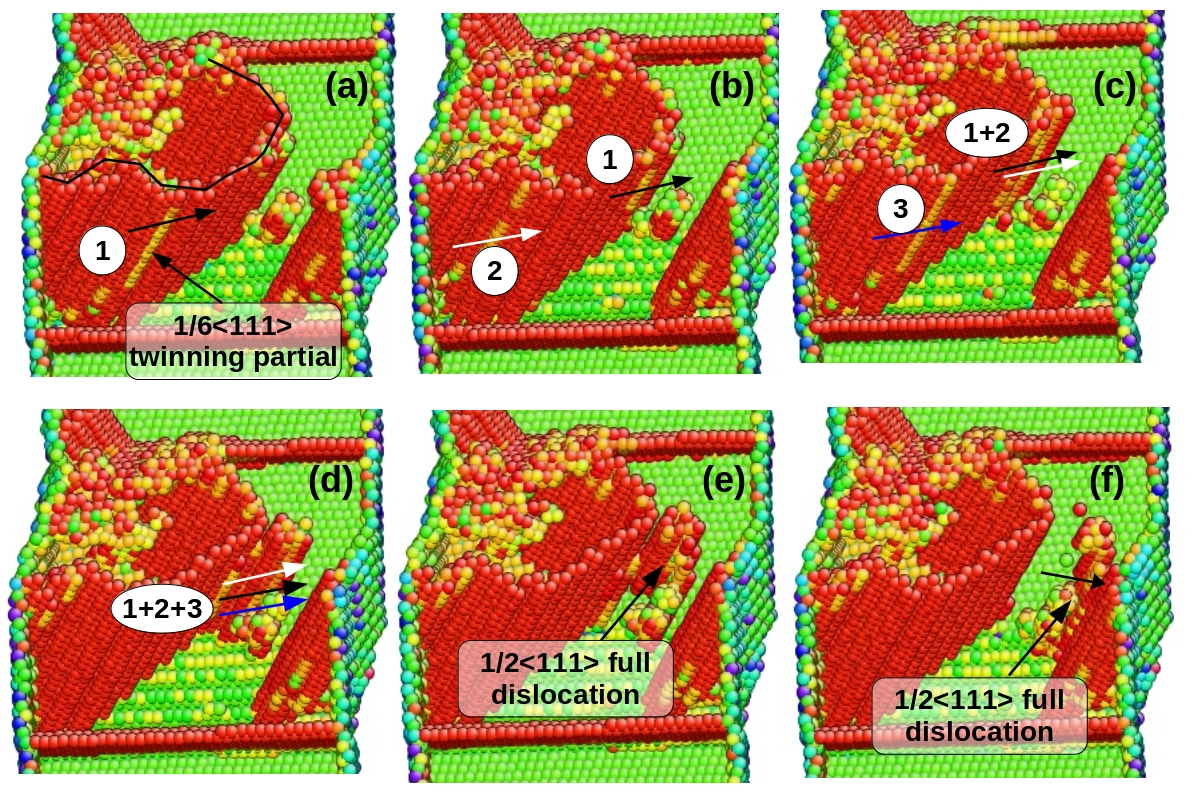}
\caption {\footnotesize Process of full dislocation nucleation from the edge of curved twin boundary during 
tensile deformation of twinned BCC Fe nanopillar. The atoms are coloured according to the centro-symmetry 
parameter \cite{CSP}. The perfect BCC atoms and the front surface are removed for clarity.}
\label{full-slip}
\end{figure}

\section{Discussion}

\subsection{Non-linear elastic deformation}

In tensile loading, all the nanopillars exhibited the non-linear elastic deformation at high elastic strains 
(Figure \ref{stress-strain-tension}), while in compressive loading only twinned nanopillars displayed non-linear 
behaviour (Figure \ref{stress-strain-compress}). The nanopillar without twin boundary under compressive loading 
showed perfectly linear elastic deformation (Figure \ref{stress-strain-compress}). In the past, many atomistic 
simulations studies have shown that the elastic deformation of various FCC and BCC metallic nanowires is non-linear 
at high elastic strains under different loading conditions \cite{Sainath-CMS16,Nonlinear}. The non-linear elastic 
deformation 
observed at high strains in the present investigation can be attributed to the nature of inter-atomic forces in 
BCC Fe nanowires \cite{Sainath-CMS16}. At small atomic distances, the inter-atomic forces vary linearly with 
inter-atomic distance and this reflects in a linear elastic deformation at small strains. However, at higher 
atomic distances, the inter-atomic force varies non-linearly and this leads to the appearance of non-linear 
elastic deformation at high strains. Due to negligible defect density and high surface effects, the defect 
nucleation in the pristine nanowires requires high elastic strains compared to their bulk counterparts. As a 
result, the non-linear elastic deformation is observed mainly in materials at nanoscale. The observed non-linearity 
in elastic deformation results as a consequence of the ability of the nanowires to undergo large elastic deformation. 
In general, all the nanopillars under tensile and compressive loadings have shown large elastic strains except 
perfect nanopillar showing comparatively small elastic strain under compressive loading.

\subsection{Effect of twin boundary spacing on yield stress}

The yield stress as a function of twin boundary spacing exhibiting contrasting behaviour under tensile and compressive 
loadings has been observed in BCC Fe nanopillars. During tensile deformation, the yield stress displays negligible 
variation with respect to twin boundary spacing (Figures \ref{stress-strain-tension} and \ref{TBS-Tension}), while a 
significant decrease in yield strength with increasing twin boundary spacing has been observed under compressive loading 
(Figures \ref{stress-strain-compress} and \ref{TBS-Compression}). The observed variations in yield stress with respect 
to twin boundary spacing under tensile and compressive loadings are in agreement with those reported in twinned FCC 
nanopillars \cite{Cao,Zhang,Hammami,Wei}. In twinned FCC nanopillars, the strengthening and negligible influence of 
twin boundary spacing was explained using the hard and soft modes of deformation \cite{modes,modes-2}. In BCC Fe 
nanopillars, this contrasting behaviour in the yield stress variations can arise mainly from the difference in yielding 
and subsequent plastic deformation mechanisms under tensile and compressive loadings. The observed deformation mechanisms 
are summarised schematically in Figure \ref{modes}. Since the deformation under tensile loading is dominated by the 
twinning partial dislocations (Figures \ref{deform-tension}-\ref{All-TB-Tension}), the glide plane and the Burgers vector 
of this partials can either have an inclination or parallel with respect to twin boundary (Figure \ref{modes}(a)). In 
case of an inclination, the repulsive force on the partial dislocations is smaller than that of full dislocations and 
as a result, the yield stress under tensile loading exhibits weaker dependence on twin boundary spacing. In other case, 
where the glide plane and the Burgers vector of partial dislocations are parallel to twin boundaries (Figure \ref{modes}(a)), 
the twin boundary spacing has a insignificant role on yield stress due to negligible repulsive force  \cite{modes,modes-2}. 
On the other hand, when the deformation is dominated by full dislocations under compressive loading (Figures \ref{deform-compress} 
and \ref{All-TB}), the slip plane and the Burgers vector of full dislocations have an inclination with respect to the 
twin boundary (Figure 
\ref{modes}(b)). This mode of deformation is known as hard mode in FCC nanopillars \cite{modes}. In this mode, the twin 
boundaries exert strong repulsive force on full dislocations and also influence the resistance to dislocation slip. The 
repulsive force on the dislocations increases upon decreasing twin boundary spacing and as a result, the yield stress 
increases with the decrease in twin boundary spacing.

\begin{figure}[h]
\centering
\includegraphics[width= 9cm]{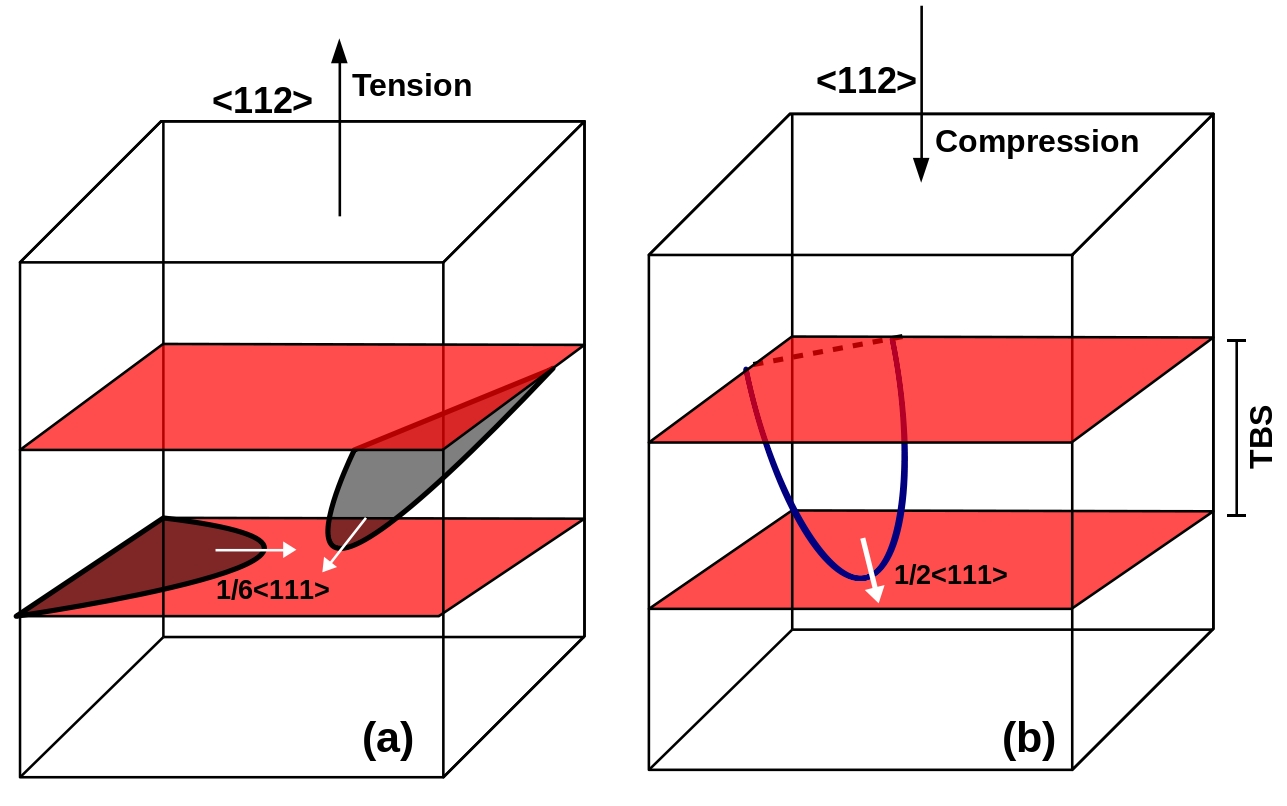}
\caption {\footnotesize Schematic of the observed deformation mechanisms under (a) tensile and (b) compressive 
deformation of BCC Fe nanopillars.}
\label{modes}
\end{figure}

The yield stress of twinned nanopillars under tensile loading is much lower than the yield stress of perfect nanopillar 
(Figure \ref{TBS-Tension}), while in compression the strength values are comparable for small spacing (i.e. nanopillars 
having three and five twin boundaries) and lower, for large spacing (Figure \ref{TBS-Compression}). This difference in 
yield stress between perfect and twinned nanopillars arises mainly from the structure of twin boundary. With Mendelev 
EAM potential, the twin boundary in BCC Fe has a displaced or isosceles structure \cite{Sainath-CMS15, Sainath-CMS16,
Scripta-16}, where the upper grain is displaced with respect to the lower grain parallel to the boundary plane by the 
vector 1/12$<$111$>$. This extra step was observed experimentally in BCC Mo \cite{GB-Mo, GB-Mo-1}, but not in BCC Fe. 
In BCC Fe, it has been reported that the twin boundary posses a reflection structure \cite{GB-Fe,GB-Fe-1}. In twinned 
nanopillars, this extra step can acts as a nucleation site for a twin embryo, while in perfect nanopillars, the twin embryo 
nucleates from the corner. Due to this extra step, the defect nucleation in the twinned nanopillars is easier than in perfect 
nanopillar. Under compressive loading, the observed low yield stress values for higher twin boundary spacing (i.e. 
nanopillars with one and two twin boundaries) can arise from the presence of such step. Further, the observed yield 
stress values for low twin boundary spacing (i.e. nanopillars with three and five twin boundaries) comparable to 
perfect nanopillar can be ascribed to the combined effects associated with repulsive force and step.

\subsection{Flow stress and tension-compression asymmetry}
Following yielding under tensile loading, the $<$112$>$ perfect nanopillar exhibits nearly constant low flow stress 
arising from the dominance of twin growth or twin boundary migration. It has been demonstrated that when the deformation 
is dominated by twinning mechanism in BCC Fe nanowires with $<$100$>$, $<$112$>$ and $<$102$>$ orientations, the twin 
growth/twin boundary migration remains a self-driven process and this results in nearly constant low flow stress 
\cite{Sainath-CMS15,Sainath-CMS16}. In general, deformation twinning spreads in the majority of gauge length in perfect 
nanopillars giving rise to large ductility. In twinned nanopillars, the gauge length portion participating during tensile 
deformation decreases with the increase in the number of twin boundaries or decrease in twin boundary spacing (Figures 
\ref{deform-tension} and \ref{All-TB-Tension}). As a result, the nanopillars containing one and two twin boundaries display 
gradual decrease in flow stress, well-defined necking and lower tensile ductility (Figures \ref{stress-strain-tension} and 
\ref{All-TB-Tension}(a)). In case of nanopillars having three and five twin boundaries, more and more localised deformation 
leads to intense necking, rapid decrease in flow stress and further reduction in tensile ductility (Figures 
\ref{stress-strain-tension} and \ref{All-TB-Tension}(b) and (c)). These results indicate that the presence of twin boundaries 
in BCC Fe nanopillars decreases the ductility compared to perfect nanopillars. In general, the twinned nanopillars exhibit 
flow stress fluctuations, which increases with increase in the number of twin boundaries particularly at low plastic strains 
(Figure \ref{stress-strain-tension}). The observed flow stress fluctuations can be attributed to dislocation activity aided 
by twin-twin interactions occurring during plastic deformation (Figure \ref{cross-slip}(a)-(e)).

In contrast to tensile loading, the deformation is completely dominated by the slip of full dislocations and deformation 
spreads throughout the nanopillar gauge length under compressive loading. The twinning mechanism has not been observed 
during compressive deformation of perfect as well as twinned nanopillars. The continuous nucleation, glide and annihilation 
of full dislocations results in large flow stress fluctuations under compressive loading in the perfect nanopillar (Figure 
\ref{stress-strain-compress}). In case of twinned nanopillars, in addition to continuous nucleation, glide and annihilation 
of full dislocations, dislocation-twin boundary interactions contribute to the observed flow stress fluctuations (Figures 
\ref{stress-strain-compress}, \ref{TB-inter-1} and \ref{TB-inter-2}). The observed tension-compression asymmetry in deformation 
mechanisms in terms of twinning under tensile loading and dislocation slip under the compressive loading is in agreement with 
those observed in $<$100$>$ BCC Fe nanowires \cite{Healy-2015,Sainath-CMS16}. In BCC nanowires, tension-compression asymmetry 
has been demonstrated for yield stress \cite{Sainath-CMS16,Greer-Mo} as well for the operating deformation mechanism 
\cite{Healy-2015,Sainath-CMS16,BCC-W-exp}. It has been shown that the tension-compression asymmetry in deformation mechanisms 
in BCC Fe nanowires results as a consequence of twinning-antitwinning sense of 1/6$<$111$>$ partial dislocations on \{112\}
planes under the opposite loading conditions \cite{Healy-2015,Sainath-CMS16,Greer-Mo}.

\subsection{Twin-twin and dislocation-twin interactions}

The plastic deformation and the associated flow stress in polycrystalline materials are mainly controlled by the interaction 
between dislocations and grain boundaries. Here, we discuss the twin-twin and dislocation-twin interactions observed during 
deformation in the twinned nanopillars. In BCC metals, the twin-twin interactions have been classified based on their common 
line of intersection \cite{Mahajan}. The observed $<$021$>$ type twin intersection is in agreement with the fact that in BCC 
metals, there are only five types of probable twin-twin intersections, namely $<$011$>$, $<$012$>$, $<$113$>$, $<$111$>$ and 
$<$135$>$ \cite{Mahajan}. In $<$021$>$ type twin intersection, the 1/2$<$111$>$ full dislocation may also originate in the 
neighbouring grain by the coalescence of 1/6$<$111$>$ twinning partial dislocations (Figure \ref{cross-slip}(d)-(e)). Along 
with full dislocations, the nucleation of twinning partials gliding along the initial twin boundary leads to the twin boundary 
migration. In agreement with the present study, Ohja et al. \cite{Ojha} characterised $<$110$>$, $<$113$>$ and $<$210$>$ types 
twin-twin intersections in BCC Fe using experiments and molecular dynamics simulations. 

The dislocation nucleation from the edge
of curved twin boundary observed in the present investigation (Figure \ref{full-slip}) is interesting, and this is in agreement 
with those obtained experimentally by Hull \cite{Hull} in BCC Fe and Wang et al. \cite{BCC-W-exp} in BCC W using MD simulations. 
In BCC metals, the twin boundary or the twin-matrix interface can have different shapes such as lenticular, flame-like structure
and complicated fine structure of serrations \cite{Reid}. The edges or the serrations can acts as a stress concentration site. 
In order to relieve this localised stress, full dislocation nucleates from the edges of the curved twin boundary and glide on 
\{110\} plane \cite{Hull}.

The dislocation-twin boundary interactions during deformation under compressive loading reveal that the dislocation can either 
directly transmits through the twin boundary without any deviation in the glide plane (Figure \ref{TB-inter-1}) or it can 
transmits to symmetrical plane in the neighbouring grain (Figure \ref{TB-inter-2}). Mrovec et al. \cite{Mrovec} examined the 
dislocation-twin boundary interactions in BCC tungsten using Finnis-Sinclair and bond order potential. Depending on inter-atomic 
potential, dislocation character and its glide plane, various reactions such as dislocation splitting into three twinning 
partials along the twin boundary, twin boundary destruction and disintegration close to absorption site, the dislocation
embedding in the twin boundary, and the direct transmission of the dislocations have been observed \cite{Mrovec}. In BCC metals, 
three \{112\} and three \{110\} planes have the same $<$111$>$ zone axis (Figure \ref{transmission}(a)). Interestingly, this 
arrangement of glide planes remains same across the twin boundary (Figure \ref{transmission}(b)). In view of this similarity 
in the slip planes, dislocation once passes through the twin boundary can either glide on symmetrical plane or can directly
comes out in the neighbouring grain without any deviation in the glide plane. The inserted atomic snapshots in Figure 
\ref{transmission}(b) shows that that the observed slip lines in Figures \ref{TB-inter-1} and \ref{TB-inter-2} are parallel 
to \{112\} planes. The geometrical 
arrangement of glide plane (Figure \ref{transmission}(b)) indicates that the observed dislocation-twin boundary interactions 
in the present study are on expected lines in BCC metals.

\begin{figure}[h]
\centering
\includegraphics[width= 12cm]{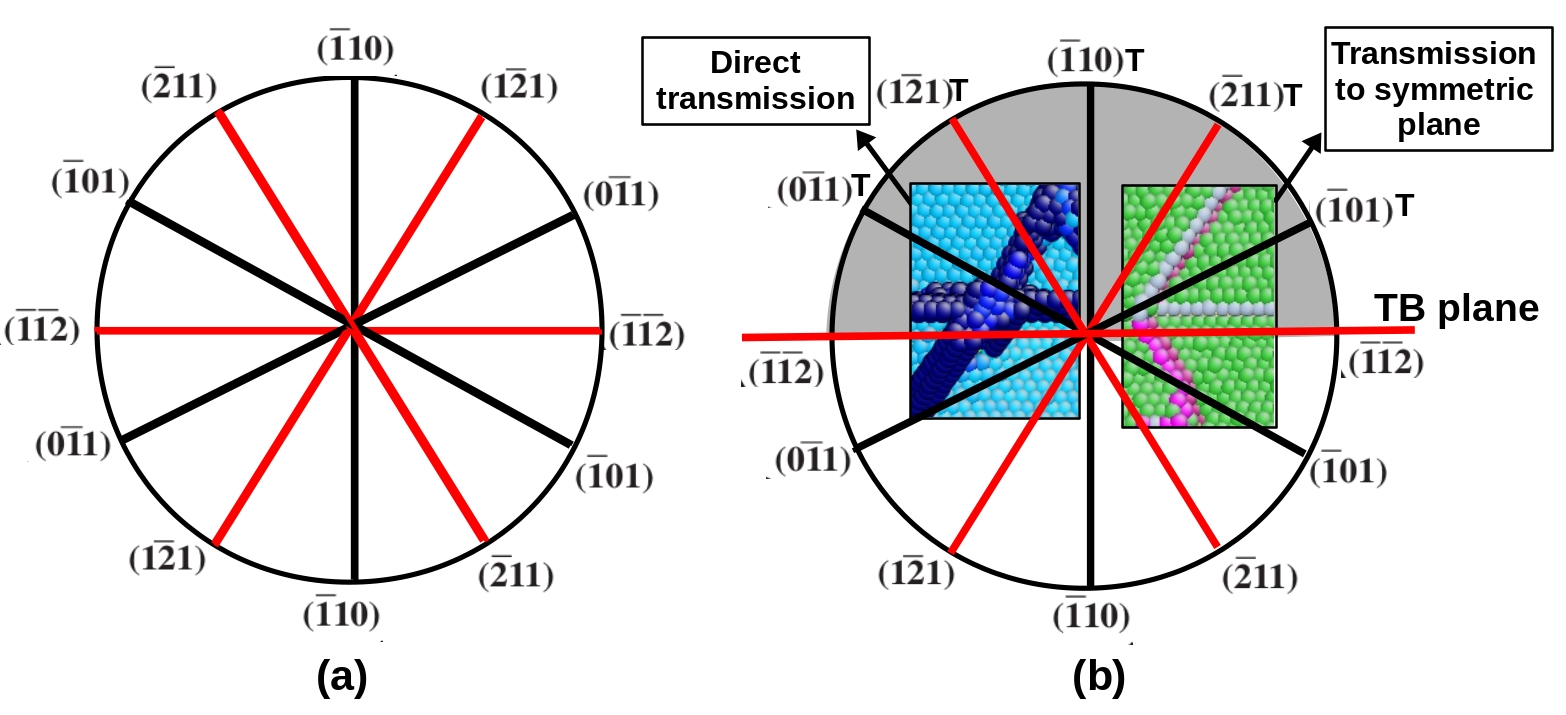}
\caption {\footnotesize The arrangement of three \{110\} and three \{112\} planes having the same $<$111$>$ zone axis in BCC 
metal in (a) perfect and (b) across the twin boundary in twinned crystals.}
\label{transmission}
\end{figure}

\subsection{Comparison with experimental data and effect of inter-atomic potential}

The most important aspect of atomistic simulations is the reliability of inter-atomic potential. Generally, the EAM 
potentials are more reliable for many FCC systems, while for BCC systems they are comparatively less accurate in 
reproducing the defect structures, slip systems, twinning behaviour and phase transition. This is mainly because 
the non-planar core of screw dislocations in BCC systems makes the slip more complex compared to FCC systems. 
Furthermore, the atomistic simulations on various BCC systems have suggested that the activated slip systems 
and the deformation mechanisms depends sensitively on inter-atomic potential employed in simulations 
\cite{Hale,Simonelli,Chaussidon}. For example in BCC Fe, Mendelev EAM potential employed in the present investigation 
shows non-degenerate core for screw dislocations and predicts \{112\} and \{110\} as glide planes \cite{Chaussidon}. 
Contrary to this, Simmonelli potential \cite{Simonelli} shows degenerate core for screw dislocations and predicts 
only \{112\} glide plane \cite{Chaussidon}. However, recent experimental results have shown that the glide is 
observed on \{110\} planes in BCC Fe \cite{Caillard}. In this context, it is more important to understand that up 
to what degree 
the predictions made in the present investigation differ with potential. Cao \cite{Cao-BCC} in his study on shape 
memory and pseudo-elasticity 
of $<$100$>$ BCC Fe nanowires has shown that under tensile deformation, the qualitative stress-strain behaviour and 
deformation by twinning doesn’t vary significantly with Mendelev \cite{Mendelev-2003} and Chamati \cite{Chamati} 
potentials. However, these two potentials predict different values of yield stress \cite{Cao-BCC}. On similar lines, 
the quantitative values of yield stress predicted in the present investigation may vary with inter-atomic potential, 
but the qualitative aspects may not change significantly although it needs further investigations.

In the present study, we have chosen Mendelev EAM potential mainly because several predictions made with this 
potential are in good agreement with either experimental observations or density functional theory calculations. 
The first and foremost is the Mendelev EAM potential predicting a non-degenerate core structure for screw dislocations 
spreading symmetrically on three \{110\} planes of $<$111$>$ zone \cite{Chaussidon}. This is in agreement with
density functional theory calculations \cite{DFT}. All other potentials including Chamati \cite{Chamati} and 
Simonelli \cite{Simonelli} potentials for BCC Fe predicts a degenerate core structure spreading asymmetrically 
on three \{110\} planes \cite{Chaussidon}. Similarly, the predictions made in the present investigation on 
deformation twinning and dislocation slip in BCC Fe nanopillars are quite close to the experimental observations 
in BCC W nanopillars \cite{BCC-W-exp}. Using in situ transmission electron microscopy in BCC W nanopillars, Wang 
et al. \cite{BCC-W-exp} predicted deformation twinning for  $<$100$>$-tension,  $<$110$>$ and $<$111$>$-compression, 
while dislocation slip is observed for  $<$112$>$-tension and compression. In agreement with this experimental 
study, the Mendelev EAM potential predicts deformation twinning for  $<$100$>$-tension and  $<$110$>$-compression
and dislocation slip for  $<$112$>$-compression in BCC Fe nanowires/nanopillars \cite{Sainath-CMS16}. However, it 
predicts twinning for  $<$112$>$-tension \cite{Sainath-CMS16}, which is in contrast with the experimental 
observations \cite{BCC-W-exp}. Similarly, the mechanism of twin nucleation and growth, and twin boundary as
a source of full dislocation nucleation observed in the present investigation using Mendelev EAM potential 
are in agreement with those observed experimentally \cite{BCC-W-exp,Hull}. Further, in agreement with experimental 
results, the MD simulation results using Mendelev EAM potential also described the various twin-twin interactions 
and twin migration stress in BCC Fe \cite{Ojha}.

\section{Conclusions}

Molecular dynamics simulations performed on twinned BCC Fe nanopillars indicated that the twin boundaries have a 
contrasting role under tensile and compressive loadings. Under tensile loading, the yield stress has been almost 
independent of twin boundary spacing, while under compressive loading, the yield stress showed strong dependence 
on twin boundary spacing. This contrasting behaviour in yield stress has been explained by repulsive force offered 
by the twin boundaries. Under tensile loading, deformation is dominated by the twin growth/twin boundary migration, 
where the initial twin boundary offers negligible repulsive force on the nucleation of twinning partials. Due to this, 
yield stress varies marginally as a function of twin boundary spacing. In addition to twinning, minor activity of 
full dislocations and twin-twin interactions of $<$021$>$ type have been observed during tensile deformation. It has 
been found that the edge of the curved twin boundary can acts as a source for the emission of full dislocations. It 
has been observed that the deformation under compressive loading is dominated by the slip of full dislocations, where 
the twin boundaries offer a strong repulsive force for the nucleation of full dislocations. This leads to the observed 
strong dependence of yield stress on twin boundary spacing. The dislocation-twin boundary interactions obtained under 
the compressive deformation of twinned nanopillars revealed that the dislocation can either directly transmit without 
any deviation in slip plane or it can transmit on symmetrical slip plane in the neighbouring grain.

}

\end{document}